\documentclass[%
 reprint,
 amsmath,amssymb,
 aps,
prl,
]{revtex4-2}

\usepackage{graphicx}
\usepackage{dcolumn}
\usepackage{bm}
\usepackage{hyperref}
\usepackage[mathlines]{lineno}
\usepackage{algorithm}
\usepackage[noend]{algpseudocode}
\usepackage{xcolor}
\usepackage{pifont}
\usepackage{tikz}
\usetikzlibrary{matrix}
\usepackage[normalem]{ulem}
\usepackage{verbatim}
\usepackage{soul}

\begin{document}
\preprint{arXiv}

\title{Symmetry-broken perturbation theory to large orders in antiferromagnetic phases}

\author{Renaud Garioud$^{1,2}$}
\email{garioudrenaud@gmail.com}
\author{Fedor \v{S}imkovic IV$^{1,2}$}
\author{Riccardo Rossi$^{3,4}$}
\author{Gabriele~Spada$^{5,6}$}
\author{Thomas Schäfer$^{7}$}
\author{Félix Werner$^{5}$}
\author{Michel Ferrero$^{1,2}$}

\affiliation{
$^1$CPHT, CNRS, Ecole Polytechnique, Institut Polytechnique de Paris, Route de Saclay, 91128 Palaiseau, France\\
$^2$Coll\`ege de France, 11 place Marcelin Berthelot, 75005 Paris, France \\
$^3$Institute of Physics, École Polytechnique Fédérale de Lausanne (EPFL), CH-1015 Lausanne, Switzerland \\
$^4$Sorbonne Universit\'e, CNRS, Laboratoire de Physique Th\'eorique de la Mati\`ere Condens\'ee, LPTMC, F-75005 Paris, France\\
$^5$Laboratoire Kastler Brossel,  \'Ecole Normale Supérieure - Université PSL, CNRS, Sorbonne Université, Collège de France, 75005 Paris, France \\
$^6$ INO-CNR BEC Center and Dipartimento di Fisica, Università di Trento, 38123 Trento, Italy \\
$^7$Max-Planck-Institut für Festkörperforschung, Heisenbergstraße 1, 70569 Stuttgart, Germany}

\date{\today}

\begin{abstract}

     We introduce a spin-symmetry-broken extension of the connected determinant algorithm [Phys. Rev. Lett. 119, 045701 (2017)]. The resulting systematic perturbative expansions around an antiferromagnetic state allow for numerically exact calculations directly inside a magnetically ordered phase. We show new precise results for the magnetic phase diagram and thermodynamics of the three-dimensional cubic Hubbard model at half-filling. With detailed computations of the order parameter in the low to intermediate-coupling regime,
    we establish the N{\'e}el  phase boundary. The critical behavior
    in its vicinity is shown to be compatible with the $O(3)$ Heisenberg universality class. By determining the evolution of the entropy with decreasing temperature through the phase transition we identify the different physical regimes at $U/t\!=\!4$.
    We provide quantitative results for several thermodynamic quantities deep
    inside the antiferromagnetic dome up to large interaction strengths and investigate the crossover between the Slater and Heisenberg regimes. 
\end{abstract}

\maketitle

 In strongly correlated materials, such as high temperature superconducting copper oxides or iron-based pnictides, the interactions between electrons
yield intricate phase diagrams, exhibiting, e.g., magnetically or charge-ordered phases, superconductivity or Mott insulating behaviors. Understanding the properties of these different phases, their interplay and driving mechanisms is one of the outstanding challenges of modern condensed matter theory.

From the theoretical point of view, one of the simplest models to investigate phase transitions is the three-dimensional cubic Hubbard model~\cite{Hubbard1963,Hubbard1964,Kanamori1963,Gutzwiller1963,Hubbard_gen_1,Hubbard_gen_2} given by the Hamiltonian
\begin{equation}
         \mathcal{\hat{H}}=-t\sum_{\langle i,j \rangle}\sum_{\sigma}\hat{c}_{i\sigma}^{\dagger}\hat{c}_{j\sigma} + U \sum_{i}\hat{n}_{i\uparrow}\hat{n}_{i\downarrow} - \mu \sum_{i\sigma}\hat{n}_{i\sigma},
\label{Hubbard}
\end{equation}
where $t$ is the hopping amplitude between nearest-neighbor sites $\left<i,j\right>$
on a cubic lattice, $U\geq 0$ the on-site Coulomb interaction, $\mu$ the chemical potential, $\hat{n}_{i\sigma}=\hat{c}_{i\sigma}^{\dagger}\hat{c}_{i\sigma}$ and $\hat{c}_{i\sigma}^{\dagger}$ creates an electron on site $i$ with spin $\sigma$.
At half-filling ($\mu=U/2$), the ground state has antiferromagnetic long-range spin order. In three dimensions this SU(2) symmetry-broken phase survives up to the N{\'e}el  temperature $T_N(U)$ above which the system becomes paramagnetic.
While there is qualitative understanding of the mechanisms that produce
the antiferromagnetic order both at weak and strong coupling, obtaining unbiased quantitative results, especially close to the phase transition
and inside the ordered phase, is still very challenging~\cite{DDMC,DGA_phase_diag,DCA_size_scaling,TUFRG,Dual_fermions,QMC,2nd_order_pert_theory,DGA_3d_2d,Real_material_3D,Rohringer2018,DGA_symmetry_breaking}.
Therefore, despite its apparent simplicity, the Hubbard model on the cubic lattice is an ideal
platform to explore the potential of new algorithms before engaging in
the study of more realistic systems. The model was realized in cold-atomic experiments on optical lattices where antiferromagnetism is under active investigation~\cite{Cold_atoms_AFM_1,Cold_atoms_AFM_2,Cold_atoms_AFM_3,Cold_atoms_AFM_4,zwierlein_spin_correl,bakr_canted_AF,bloch_magnetic_polaron,kohl_bilayer,bloch_chains_haldane_phase}. 

The main challenge for theoretical approaches based on finite size lattices is to properly account for the increasing correlation length in the vicinity of a second order phase transition, and, as such, to extrapolate to the thermodynamic limit. In that respect, the diagrammatic Monte Carlo approach~\cite{prokofiev_svistunov_1998,VanHoucke1short,KozikVanHouckeEPL} is very promising
as it offers the possibility to investigate a system directly in the
thermodynamic limit. The method stochastically computes the coefficients $a_{k}$ appearing in the perturbative expansion in $U$ of a physical observable,
$\mathcal{A}(U) = \sum_{k}a_{k}U^{k}$ in the simplest formulation. The computational cost rapidly
increases with increasing perturbation orders and only
 so many coefficients can be computed before the statistical variance becomes overwhelming. Nevertheless, important improvements~\cite{CDet_first,  CDet_Polynomial_complexity} make it now possible to reach perturbation orders as large as $10-12$. In the context of the repulsive Hubbard model, diagrammatic Monte Carlo has already been successfully applied to
non-perturbative regimes in the two-dimensional square lattice~\cite{extended_crossover_2d,monster,Spin_and_charge_2D,Entropy_2d, benchmark_leblanc,shift_1,KozikVanHouckeEPL,rossi2020renormalized,benchmark_leblanc,shift_3,Conor,SimkovicPG}.

In the usual formulation, the perturbation series is constructed starting from
the non-interacting ($U=0$) SU(2)-symmetric solution of Eq.~\eqref{Hubbard}. This allows to
obtain results for the interacting system in its paramagnetic regime. As the phase transition to the antiferromagnetic state is approached, however, the resummation of the
series becomes increasingly difficult. The reason is that a second-order
phase transition happening at $U = U_c$ is accompanied by a singularity in the complex-$U$ plane for observables $\mathcal{A}(U)$ that show a non-analyticity at $U_c$.
Consequently, investigating the antiferromagnetic transition in the cubic Hubbard model can only be done from temperatures above and not
too close to the N{\'e}el  temperature $T_N$. Very recently, the spin structure factor perturbation series has been computed this way in the paramagnetic phase of the cubic Hubbard model~\cite{Conor}.
Assuming the critical behavior in the vicinity of the phase transition, the authors were able to accurately compute
$T_N$ in the weak-to-intermediate coupling regime both at half-filling and
at finite doping. This approach is however not able to address the properties of the model inside the ordered phase.

In this Letter, we take a complementary approach and compute the perturbation series for physical observables within the antiferromagnetic phase of the cubic half-filled Hubbard model. We show that our broken-symmetry approach to perturbative expansions is a powerful tool for studying magnetically ordered phases and phase transitions. Our results are obtained directly in the thermodynamic limit and, thus, do not involve any finite size scaling.  We document the vanishing of the magnetic order parameter at $T_N$ and the corresponding  critical exponent $\beta$ and report and discuss the behavior of the double occupancy, grand potential and entropy across the phase transition and inside the ordered phase. 

{\it Method.} The possibility to construct symmetry-broken perturbation series comes from a flexibility in the choice of the starting point around which the perturbation is expanded. This freedom has been extensively applied to diagrammatic Monte Carlo computations in the nonmagnetic phase to improve the convergence properties of the series \cite{shift_1,shift_2,rossi2020renormalized,simkovic2020efficient,shift_3,Homotopic_action,LeBlanc_EG_real_freq,RossiEOS,KunHauleEG,ShiftedAction,IgorHaldane,HuangPyro}. Very recently, it has been used to construct a perturbation theory around a BCS state and inside the superconducting phase of the attractive Hubbard model~\cite{Spada}. Here, we follow similar steps and introduce the modified Hamiltonian
\begin{equation}
\begin{aligned}
\hat{\mathcal{H}}_{\xi} &= -t\sum_{\langle i,j \rangle} \sum_{\sigma}\hat{c}_{i\sigma}^{\dagger}\hat{c}_{j\sigma}
- \xi \frac{U}{2} \sum_{i\sigma}\hat{n}_{i\sigma} \\
& \quad + (1-\xi) h \sum_{i} p_i \hat{S}^z_i
+ \xi U \sum_{i}\hat{n}_{i\uparrow}\hat{n}_{i\downarrow},
\end{aligned}
\label{HubbardField}
\end{equation}
where $\hat{S}^z_i\!=\!(\hat{n}_{i\uparrow} - \hat{n}_{i\downarrow})/2$ and $p_i\!=\!\pm 1$ depending on whether $i$ belongs to one or the other sub-lattice of the bipartite cubic lattice. Observables are expressed as perturbation series in $\xi$ and physical results are recovered for $\xi = 1$ where both Hamiltonians become equivalent, $\hat{\mathcal{H}}_{\xi=1} = \hat{\mathcal{H}}$.
The perturbation series in $\xi$ is built around a state that breaks the SU(2) spin rotation symmetry of the original Hamiltonian. Indeed, $\hat{\mathcal{H}}_{\xi=0}$ describes free electrons in a staggered external magnetic field $h$. Because this state breaks the symmetry from the start, the perturbation series can describe a magnetically ordered phase without the need of undergoing a phase transition. Accordingly, singularities in the complex-$\xi$ plane associated to the phase transition are avoided.

We compute the coefficients of the perturbation series with the CDet~\cite{CDet_first} algorithm using a rejection-free many-configuration Monte Carlo~\cite{Heatbath_MC} as well as a fast principal minor algorithm~\cite{Principal_minor_1,Principal_minor_2} to improve the speed of the determinant calculations.
The series are evaluated with different resummation techniques~\cite{Dlog,Resummation_tech_Fedor} that serve as a basis to determine the error bars of our results, see Supplementary Material~\cite{Supplemental}.
While the diagrammatic expansion can be formulated directly in the thermodynamic limit, in practice, we use a system with $L^3 = 20^3$ sites for our computations. We have carefully checked that this is large enough to avoid finite-size effects, even in the vicinity of the phase transition, as discussed in the Supplementary Material~\cite{Supplemental}.
In the following, we will denote this spin symmetry-broken algorithm by CDet(AF).

In the Hamiltonian of Eq.~\eqref{HubbardField}, the  field $h$ can be chosen arbitrarily and different choices for $h$ define different series. In order to obtain the best convergence and to cross-check different results, we have computed several values in the range $0 \le h \le h_\mathrm{MF}$, where $h_\mathrm{MF}$ is the effective field found in the mean-field solution of Eq.~\eqref{Hubbard}. In the following, we will parameterize $h = \alpha h_\mathrm{MF}$ with $0 \le \alpha \le 1$. Note that when $\alpha = 0$, the perturbation series is the usual expansion limited to the paramagnetic regime.

\begin{figure}[b]
\centering
\includegraphics[width=0.56\textwidth]{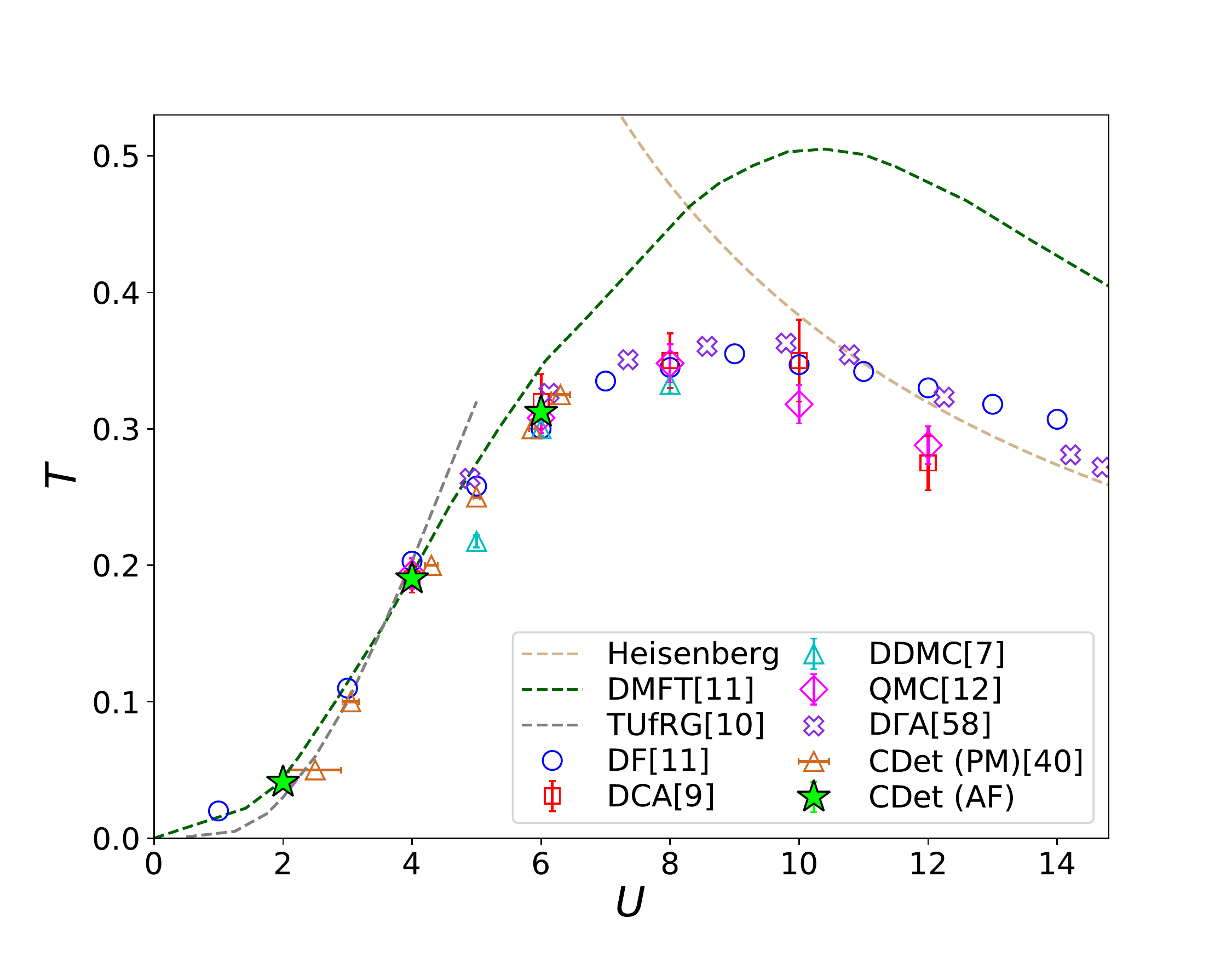}

\caption{Comparison of the N{\'e}el  temperature $T_{N}(U)$ (lime stars), obtained with the symmetry broken CDet(AF), with other numerical methods. References for the numerical methods data are indicated in the legend. }
\label{Phase_diag}
\end{figure}

For our analysis we compute the double occupancy $D\!=\!\left<\hat{n}_{i\uparrow}\hat{n}_{i\downarrow}\right>\!=\!E_\text{pot}/U$, the staggered magnetization $m=\left<\hat{n}_{i\uparrow}-\hat{n}_{i\downarrow}\right>$ (which is the order parameter for the N{\'e}el  phase transition) and the grand potential per lattice site $-\Omega/L^{3}=P$ where $L$ is the linear system size of the cubic lattice, and $P$ the thermodynamic pressure. The grand potential computations enable us to determine the entropy density and magnetization through
\begin{equation}
\label{deriv_expr}
   s=-\frac{\partial \Omega}{L^{3}\partial T}  \qquad  m=-\left.\frac{\partial \Omega}{L^{3}\partial H_\mathrm{ext}} \right|_{H_\mathrm{ext}=0},
\end{equation}
 where $H_\mathrm{ext}$ is an external Zeeman staggered field whose sign alternates on neighboring sites in the form of an additional term to the Hubbard Hamiltonian Eq.~(\ref{Hubbard}): $H_\mathrm{ext}\sum_{i} p_i \hat{S}^z_i$.  All energies are expressed in units of the hopping amplitude $t\!=\!1$.

{\it Phase diagram and universality class.} We start our study by determining the N{\'e}el  temperature for different values of the interaction in order to establish  the magnetic phase diagram of the system.

In Fig.~\ref{Phase_diag} we compare our values for the critical temperature $T_{N}(U)$ from CDet(AF) against numerous other numerical methods \cite{DDMC,Stobbe2022,Conor,Dual_fermions,QMC,DCA_size_scaling,TUFRG}. The N{\'e}el  temperature is expected to increase with increasing interaction at small $U$ since the transition is driven by the Slater mechanism \cite{Slatter} and reaches a maximum in the intermediate coupling around $U\simeq 6\!-\!10$, before decreasing like $T_{N}\simeq 0.946J$ \cite{Heisenber_Tc} in the high-$U$ Heisenberg limit, where $J=4t^{2}/U$ is the super-exchange coupling. We have been able to determine the critical temperature up to an intermediate coupling strength of $U=6$. For $U>6$, regarding the magnetization, we experience increased difficulty in resumming our perturbation series and loss of Monte Carlo accuracy in the critical region close to the phase transition.

\begin{figure}[t!]
\centering
\includegraphics[width=0.5\textwidth]{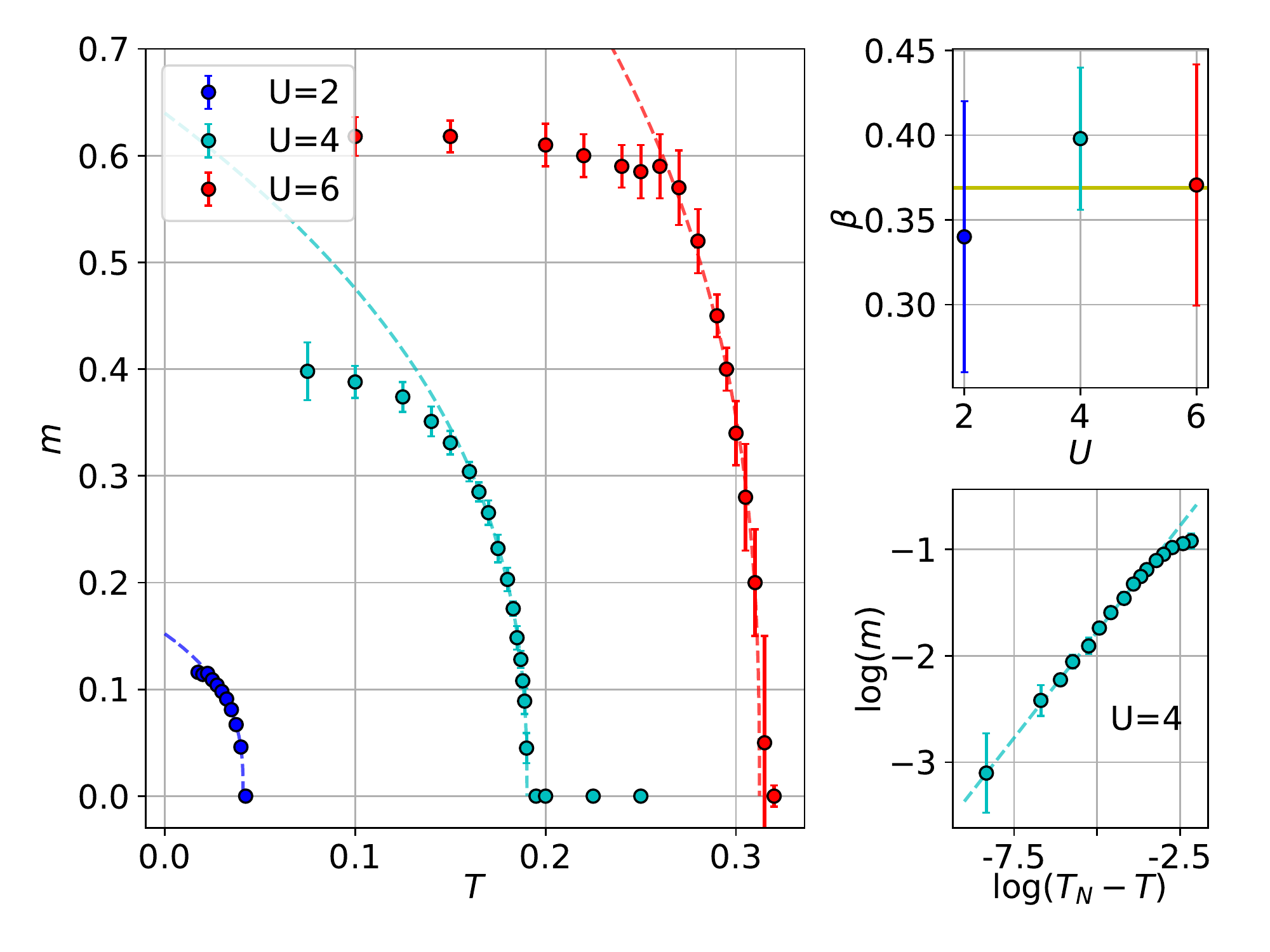}

\caption{Magnetization and critical behavior - Left panel: Magnetization $m$ as a function of temperature $T$ for three different values of the interaction $U$. The dashed curves represent the critical behavior as determined from our data close to the critical temperature fitted with the formula: $m(T)\simeq a(T_{N}-T)^{\beta}$. Top right panel: Critical exponent obtained from the three magnetization curves. The gold horizontal band corresponds to the theoretically predicted value of $\beta$ for the $O(3)$ Heisenberg universality class in \cite{Crit_exp}. Lower right panel: magnetization as a function of $T_{N}-T$ at $U=4$ on a log-log scale. $T_{N}$ is determined with the critical behavior fit from the left panel. The dashed line corresponds to the fitting curve on the left panel. 
}
\label{m_T}
\end{figure}

\begin{figure}[t!]
\centering
\includegraphics[width=0.5\textwidth, height=0.35\textwidth,keepaspectratio,]{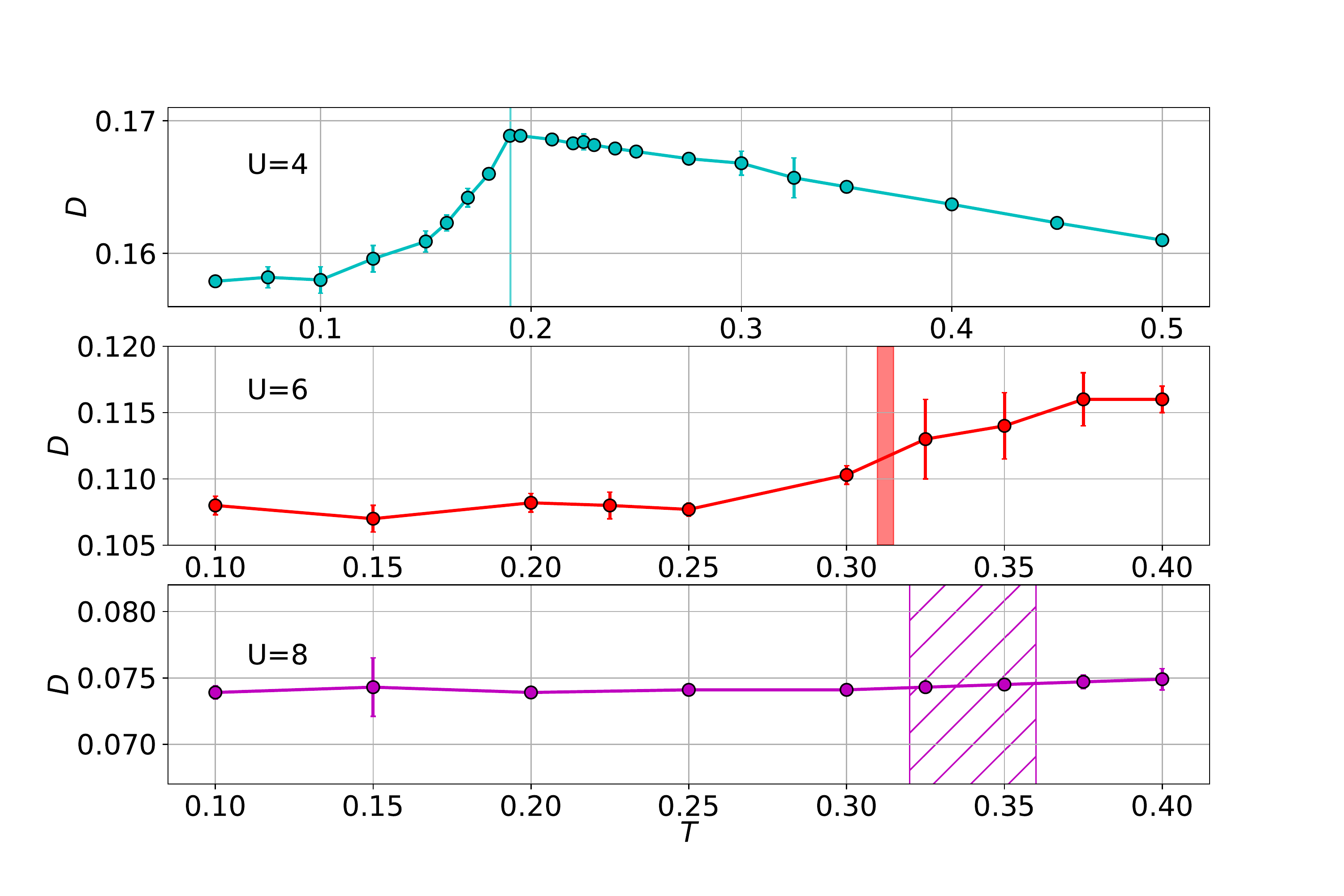}

\caption{Double occupancy $D$ as a function of temperature for three different values of the interaction $U$. The  vertical bands at $U=4$ and $U=6$ correspond to the estimate of the N{\'e}el  temperature from our study, while the hashed area at $U=8$ is an estimate of the N{\'e}el  transition from other numerical methods displayed in Fig.~\ref{Phase_diag}.}
\label{D_T}
\end{figure}

\begin{figure}[t!]
\centering
\includegraphics[width=0.5
\textwidth]{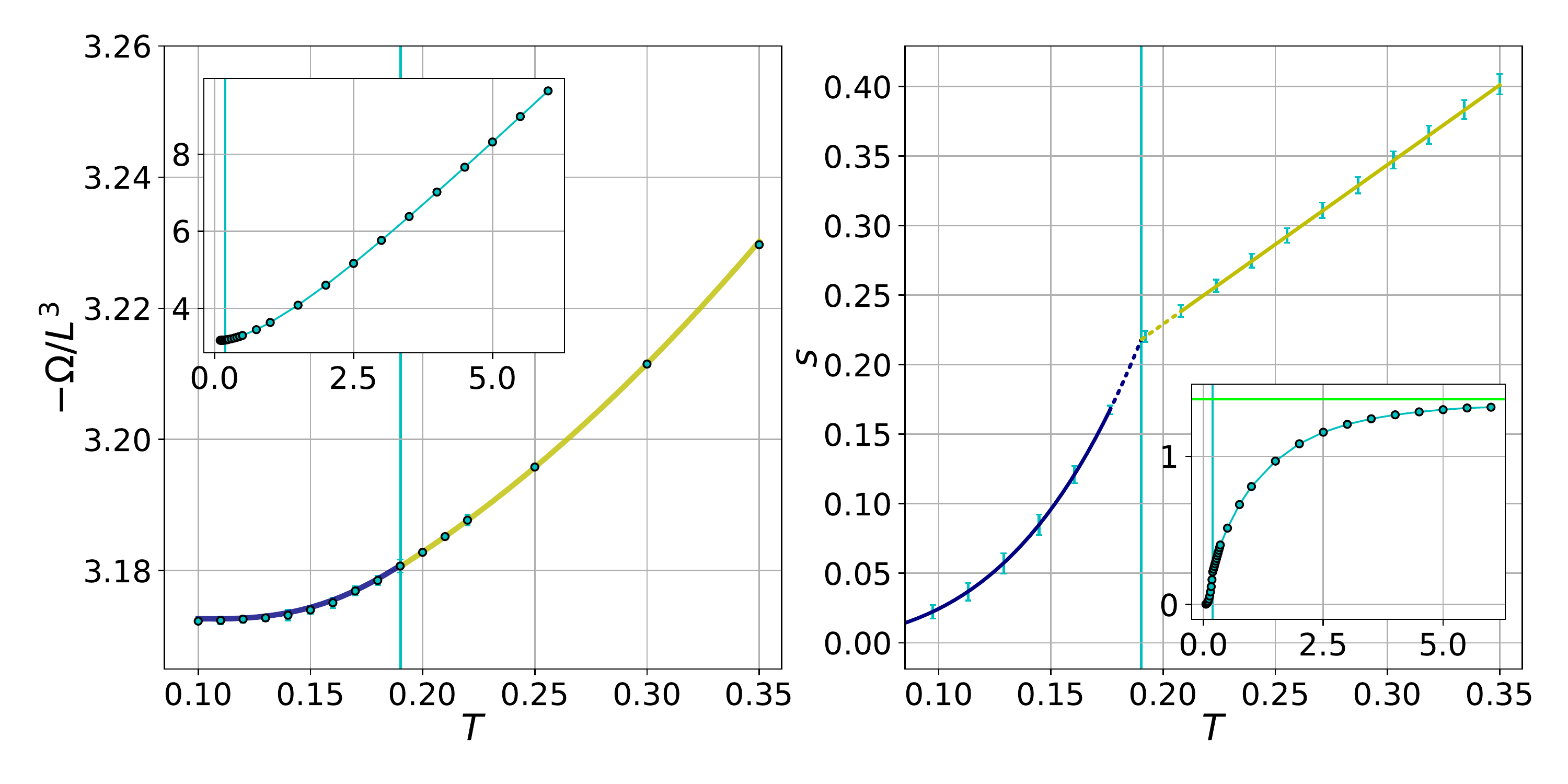}

\caption{Left panel:  Grand potential density $-\Omega/L^{3}$ as a function of temperature $T$ at interaction $U=4$. When not visible the error bar is smaller than the markers. Right panel: Entropy density $s$ as a function of temperature $T$ obtained as derivative of the grand potential fitting curves (see text). The cyan error bars give the error on the entropy curve. We do not have enough data close to the Néel temperature to resolve the critical behaviour of the entropy, and the entropy curve is dashed in this region. The insets are the same plots on a larger temperature range. The lime horizontal line indicates the high temperature limit $s=\log (4)$. On both panels the vertical lines correspond to the value of the N{\'e}el  temperature obtained in Fig.~\ref{m_T}.}
\label{P_T}
\end{figure}

The values of the N{\'e}el temperature displayed in Fig.~\ref{Phase_diag} are obtained from the computation of the magnetization as a function of temperature $m(T)$, which we show in Fig.~\ref{m_T}. The order parameter $m$ indicates the phase transition by assuming a non-zero value when decreasing the temperature : $T_{N}(U\!=\!2)\!=\!0.0425(25)$,  $T_{N}(U\!=\!4)\!=\!0.1925(25)$ and $T_{N}(U\!=\!6)\!=\!0.3125(25)$. Thanks to our high precision data, we manage to compute directly the $\beta$ critical exponent from the critical behaviour $m(T)\simeq a(T_{N}-T)^{\beta}$. The obtained values for the critical exponent Fig.~\ref{m_T} (top right) compare remarkably well to the literature values for the $O(3)$ Heisenberg universality class \cite{Crit_exp,DGA_phase_diag,Twist,Dual_fermions}. They establish the first direct computations of the $\beta$ critical exponent on a fermionic lattice and in the thermodynamic limit. As shown in Fig.~\ref{Phase_diag}, the values that we obtain for the N{\'e}el  temperature compare well with paramagnetic DiagMC \cite{Conor} and DCA extrapolated to infinite cluster size \cite{DCA_size_scaling}, as well as to the recently improved dynamical vertex approximation D$\Gamma$A~\cite{Stobbe2022}, but are out of the error bounds obtained by finite-size scaling of $L\leq 10$ DDMC data~\cite{DDMC}. The critical region, defined as the temperature range $T\in [T_N-\delta T, T_N]$ where $m(T)= a(T_N -T)^\beta$ is a good fit to our data, is of the order of $\delta T \simeq 0.025$ for $U\geq 4$. The magnetization and the other thermodynamic quantities (see Figs~\ref{D_T} and \ref{P_T}) only have a variation
in a temperature interval $\delta T \simeq 0.1$ below $T_N$ before they essentially saturate to their low temperature value. This interval does not seem to expand when increasing the interaction and hence the N{\'e}el  temperature for $U\geq 4$.

{\it Double occupancy.} The signatures of the phase transition can also be read from the double occupancy, shown in Fig.~\ref{D_T}. At $U\!=\!4$, we observe a singularity in the double occupancy at a temperature in good agreement with the value of the N{\'e}el  temperature determined in Fig.~\ref{m_T}. At this value of the interaction, the double occupancy increases with decreasing temperature in the normal phase because of the Pomeranchuk effect~\cite{Pomeranchuk,Pomeranchuk1,Pomeranchuk2,Pomeranchuk3,Double_occupancy_entropy}. It decreases in the antiferromagnetic phase which is consistent with the Slater mechanism expected at small interaction: The ordered phase is stabilized because of a gain in potential energy $E_{pot}\!=\!UD$ and, hence, a lowering of double occupancy at fixed interaction. At higher values of $U$ the double occupancy curve flattens, and within our accuracy,  we are not able to document the non-analyticity of the double occupancy at the N{\'e}el  temperature. We do not observe significant changes of the double occupancy around the N{\'e}el  temperature at $U\!=\!8$ within the $10^{-2}$ relative accuracy of our computation. Further work with better sensitivity or studying the kinetic energy would be needed to clearly document the change from a Slater to a Heisenberg regime with a kinetic-energy driven phase transition, as was done in DMFT and extensions thereof in~\cite{Taranto2012,Fratino,Harmonic_trap_Neel}.

{\it Grand potential.} The grand potential at $U\!=\!4$ is displayed in Fig.~\ref{P_T}. In order to evaluate the entropy density from Eq.~(\ref{deriv_expr}) we suppose a polynomial behavior of the grand potential with temperature. Since $\Omega(T)-\Omega(T=0)\propto T^{4}$ for $T\rightarrow0$, we fit the $T<T_{N}$ data with the expression $-\Omega(T)= -\Omega(T\!=\!0)+aT^{4}+bT^{5}+cT^{6}$ (cyan curve). At $T>T_{N}$ we expect a quadratic behavior in the degenerate Fermi liquid regime. The data is well fitted by the expression $-\Omega/L^{3}(T)= d+eT^{2}$ (yellow curve). 
We impose continuity up to first order derivative at $T=T_{N}$. At higher temperatures $T\geq 0.4$ the grand potential becomes almost linear in temperature $-\Omega(T)\simeq \log(4)T$. The entropy density is then extracted with a finite difference scheme. These different behaviors of the grand potential lead to different physical regimes for the evolution of the entropy density with temperature. In the AF phase the entropy density varies as $s\propto T^{3}$ at small temperatures. For temperatures just above the transition $T\in [T_{N},0.35]$ the entropy density increases linearly with the temperature which is a signature of a metallic behaviour of the system in this part of the phase diagram. At higher temperatures of the order of the interaction $T\sim U =4$ the entropy density saturates to $s(T\rightarrow +\infty)= \log(4)$.

{\it Magnetically saturated regime at low T.} We are now interested in the low temperature properties of the system where the magnetization has reached saturation. We have observed earlier that the magnetization only changes significantly in a shell of size $\delta T \sim 0.1$ below the N{\'e}el temperature, so that the region with saturated magnetization represents an important part of the antiferromagnetic dome.

Direct computations of the magnetization become problematic for $U\!>\!6$ because the associated series are difficult to resum. At small temperature it turns out to be more practical to compute the grand potential density and extract the magnetization as its variation with the external field as stated in Eq.~\eqref{deriv_expr}. More details, and the associated computations are shown in \cite{Supplemental}.

\begin{figure}[t!]
\centering
\includegraphics[width=0.5\textwidth]{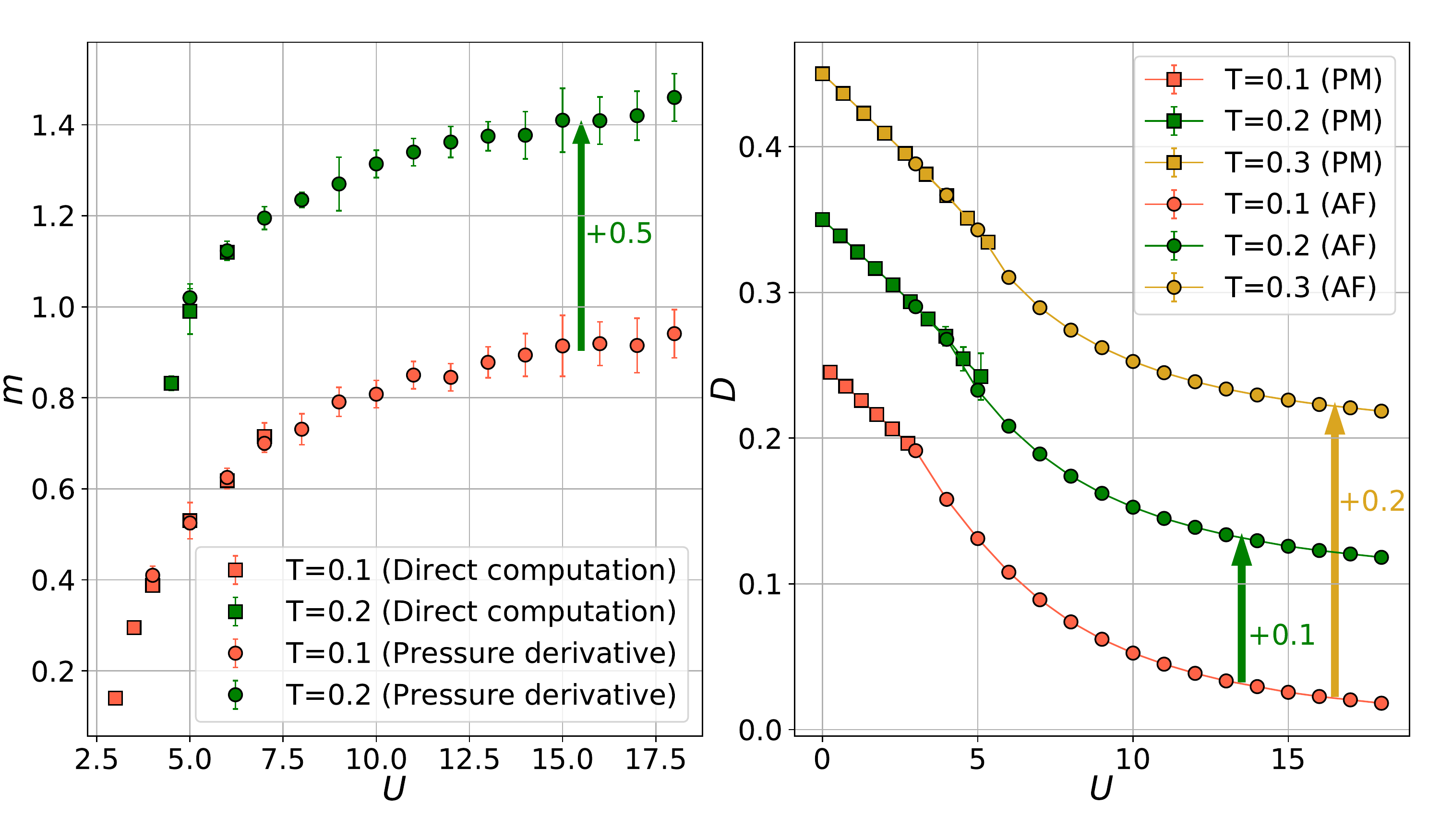}

\caption{Left panel: Magnetization as a function of the interaction $U$ for two different values of the temperature. For better visibility the magnetization at $T=0.2$ is shifted by $+0.5$ . The square markers are obtained through direct computation of the order parameter like in Fig.~\ref{m_T}. The round markers are obtained by numerically differentiating the grand potential density at an external field $H_\mathrm{ext}\rightarrow 0$, see Eq.~\eqref{deriv_expr}. Right panel: Double occupancy $D$ as a function of the interaction $U$ for three different values of the temperature. For better visibility the data at $T=0.2$ ($T=0.3$) is shifted by $+0.1$ ($+0.2$). The square markers are in the normal phase and are obtained with the paramagnetic (PM) CDet algorithm (no symmetry-breaking at $\alpha=0$). The round markers are obtained with the antiferromagnetic symmetry-broken CDet with $\alpha\neq 0$ (AF).}
\label{D_m_U}
\end{figure}

The directly computed magnetization compares well with differentiating the grand potential as shown in Fig.~\ref{D_m_U}. For $U\geq 6$ we observe no difference between the $T=0.1$ and $T=0.2$ curves which shows that the magnetization is already saturated at its zero temperature value. The magnetization will eventually have a maximum with respect to $U$, but this must happen for values of $U>18$.

The variations of the double occupancy with the interaction at low temperatures are shown in Fig.~\ref{D_m_U}. In the normal phase the double occupancy is decreasing quasi-linearly with the interaction. In the vicinity of the phase transition we observe good agreement between results for the paramagnetic and symmetry-broken computations. As expected, at the transition $U_{c}$ we observe a singularity in the double occupancy, and these results can be used to estimate the value of the critical interaction at fixed temperature. The double occupancy decreases faster with increasing interaction when entering the AF phase which is consistent with the Slater mechanism at the transition for values of the critical interaction $U_{c}<6$. In the antiferromagnetic phase the double occupancy is a convex function of the interaction which decays slowly to zero at infinite interaction. At $U> 7$ we cannot distinguish between the different temperatures within our accuracy as expected from Fig.~\ref{D_T}.

{\it Conclusions.} To conclude, we have applied the new algorithmic developments of the symmetry-broken  CDet approach to produce the first high order diagrammatic computations inside an antiferromagnetic phase and directly in the thermodynamic limit. We have provided a quantitative description of the antiferromagnetic phase of the cubic half-filled Hubbard model. After determining the critical behavior of the system and its phase diagram we have reported resummed results at small temperatures deep inside the antiferromagnetic dome up to high interactions $U=18$. We have shown that diagrammatic Monte Carlo is a powerful tool to study the physics of ordered systems with no need for an embedding scheme or system size extrapolation. 
A more advanced, non-linear chemical-potential shift combined with other CDet extensions \cite{shift_3} may lead to further improvements for describing the critical behaviour in the strong-coupling Heisenberg part of the antiferromagnetic dome. This symmetry-broken expansion could be applied to incommensurate orders in the doped regime, similarly to what was done in the normal phase \cite{Conor} or with embedding methods \cite{Twist}.
Another interesting possibility would be to extend our study by including an anisotropic hopping term $t_\mathrm{perp}<t$ in the $z$-direction (similarly to what was done in~\cite{Dare1996}) in order to investigate
how the magnetic properties are modified as the two-dimensional limit is approached.
This application would be especially relevant for the physics of cuprate superconductors.

\acknowledgments
The authors are grateful to  A.~Georges, A.J.~Kim, E.~Kozik, C.~Lenihan, G.~Rohringer, and J.~Stobbe for valuable discussions. This work was granted access to the HPC resources of TGCC and IDRIS under the allocations A0110510609 attributed by GENCI (Grand Equipement National de Calcul Intensif). This work has been supported by the Simons Foundation within the Many Electron Collaboration framework. High Performance Computing resources were provided by the IT support team of CPHT laboratory.

\bibliographystyle{unsrturl}
\bibliography{references.bib}

\end{document}


\preprint{arXiv}

\title{Supplemental material for ``Symmetry-broken perturbation theory to large orders in antiferromagnetic phases''}

\maketitle
\beginsupplement
\section{Tuning the shift ${\textstyle \alpha}$} 

We detail in this section the effect of the choice of the shift parameter on the series behaviour and discuss our method for selecting the optimal value of the $\alpha$ parameter. The expression of the perturbation Hamiltonian $\hat{\mathcal{H}}_{\xi}$  is given in Eq.~1 with $\xi$ the expansion parameter. We parameterize  $h = \alpha h_\mathrm{MF}$ with $h_{MF}$ the effective field which corresponds to the mean-field solution at interaction $U$ and temperature $T$. The free parameter $\alpha$ modifies the intensity of the staggered field in the non-perturbed system. There is no limitation prohibiting choices of $\alpha>1$, however, it was not beneficial in our studies and we limit ourselves to values of the shift $\alpha \in [0,1]$. The value $\alpha=1$ leads to a non-perturbed Hamiltonian $\hat{\mathcal{H}}_{\xi=0}$ corresponding to the antiferromagnetic mean-field Hamiltonian for which the Hartree insertions are canceled in the diagrammatic expansion of an observable $\mathcal{A}$.

\begin{figure}[H]
\centering
\includegraphics[width=0.5\textwidth]{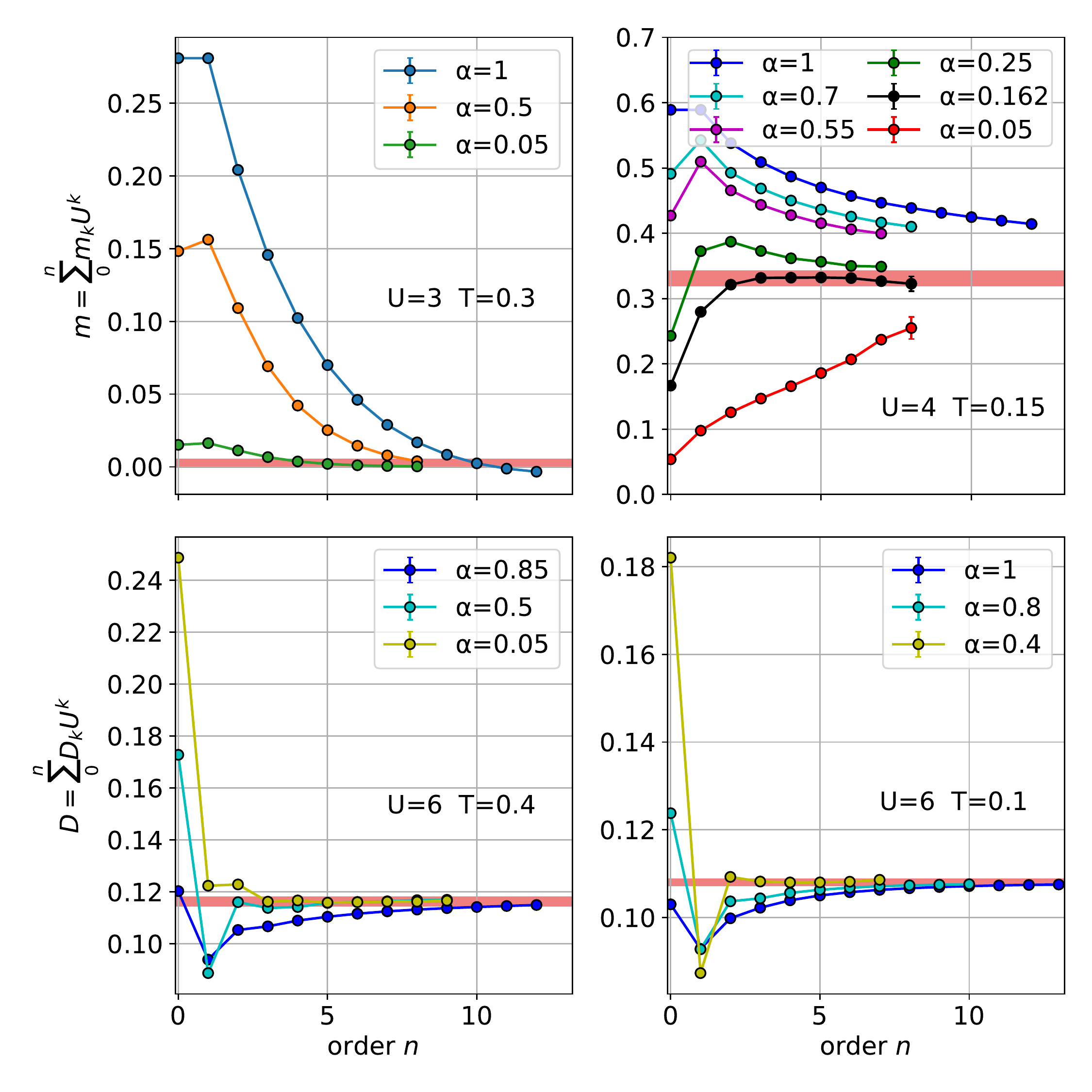} 

\caption{Partial sums of magnetization $m$ and double-occupancy $D$ for different values of the interaction $U$ and temperature $T$ as a function of the maximum order $n$. The partial sums are highly dependent on the value of the shift $\alpha$. The red colored band corresponds to the result with error after resummation.}
\label{Fig_alpha}
\end{figure}

The choice of the value of the $\alpha$ parameter changes the non-interacting Hamiltonian $\mathcal{\hat{H}}_{0}$ and therefore modifies the series behavior and its convergence properties. We illustrate in Fig.~\ref{Fig_alpha} the effect of the value of the shift parameter on the shape of the partial sums for different observables. The value of the $\alpha$ parameter leading to the best convergence of the partial sum depends on the parameters $T$ and $U$ of the computation and on the computed observable. If the best convergence is usually given by shifts in the range $\alpha \in [0.7 - 1]$, it is helpful (especially for the magnetization close to the Néel temperature) to calibrate its value in order to obtain a quickly converging partial sum. At $U=4$, $T=0.15$ we find a `magical shift' $\alpha=0.162$ for which the convergence of the series is optimal, leading to an accurate result. The maximum expansion order we are able to reach with good accuracy depends strongly on the value of the shift and is higher for $\alpha$ close to 1, which can be explained by the cancellation of Hartree insertion which reduces the MC variance.

\begin{figure}[H]
\centering
\includegraphics[width=0.5\textwidth]{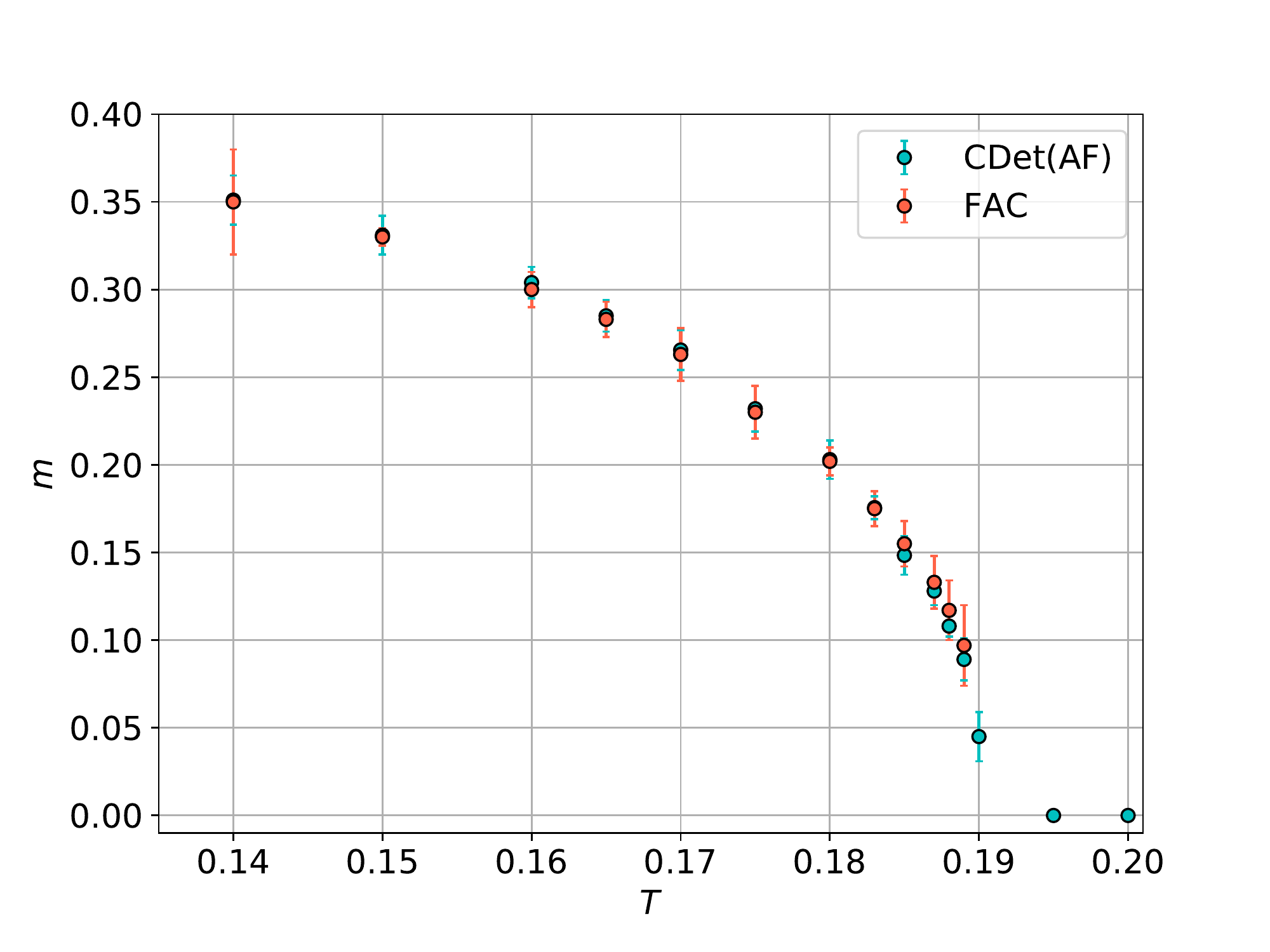} 

\caption{Magnetization as a function of temperature at interaction $U=4$ for different resummation procedures. The CDet(AF) data corresponds to the resummation procedure used in this work. See the main text for a description of the FAC resummation procedure.}
\label{Fig_mag}
\end{figure}

We employ different resummation techniques in order to extrapolate the computed series to infinite orders. We use Padé and D-log Padé approximants \cite{Resummation_tech_Fedor,Dlog} cross checked with an exponential fitting of the expansion coefficient with respect to the order to establish a controlled result. In the end the error on an observable is determined through error propagation of the MC error on the expansion orders, after comparing the results from different resummations techniques, for at least two different values of the shift parameter $\alpha$. We give in the Tables \ref{tab:D_alpha} \ref{tab:m_alpha} \ref{tab:P_alpha}, the list of all the shifts $\alpha$ used for the purpose of this work in order to obtain controlled estimates of the double-occupancy, magnetization and grand potential density. For sets of parameters corresponding to the paramagnetic phase we also rely on non symmetry-broken expansions ($\alpha=0)$. For completeness we show in Fig.~\ref{Fig_mag} a comparison of the resummation procedure which is used in this work (CDet(AF)) with the Fastest Apparent Convergence principle
(FAC) resummation procedure used and described in details in \cite{Felix}. The FAC procedure is based on optimizing the convergence of the expansion series with respect to the shift parameter $\alpha$ according to the principle of minimal sensitivity \cite{FAC1,FAC2,FAC3}. Both resummation procedures give similar results with comparable errors.  

\section{System-size effects}

\begin{figure}[H]
\centering
\includegraphics[width=0.49\textwidth]{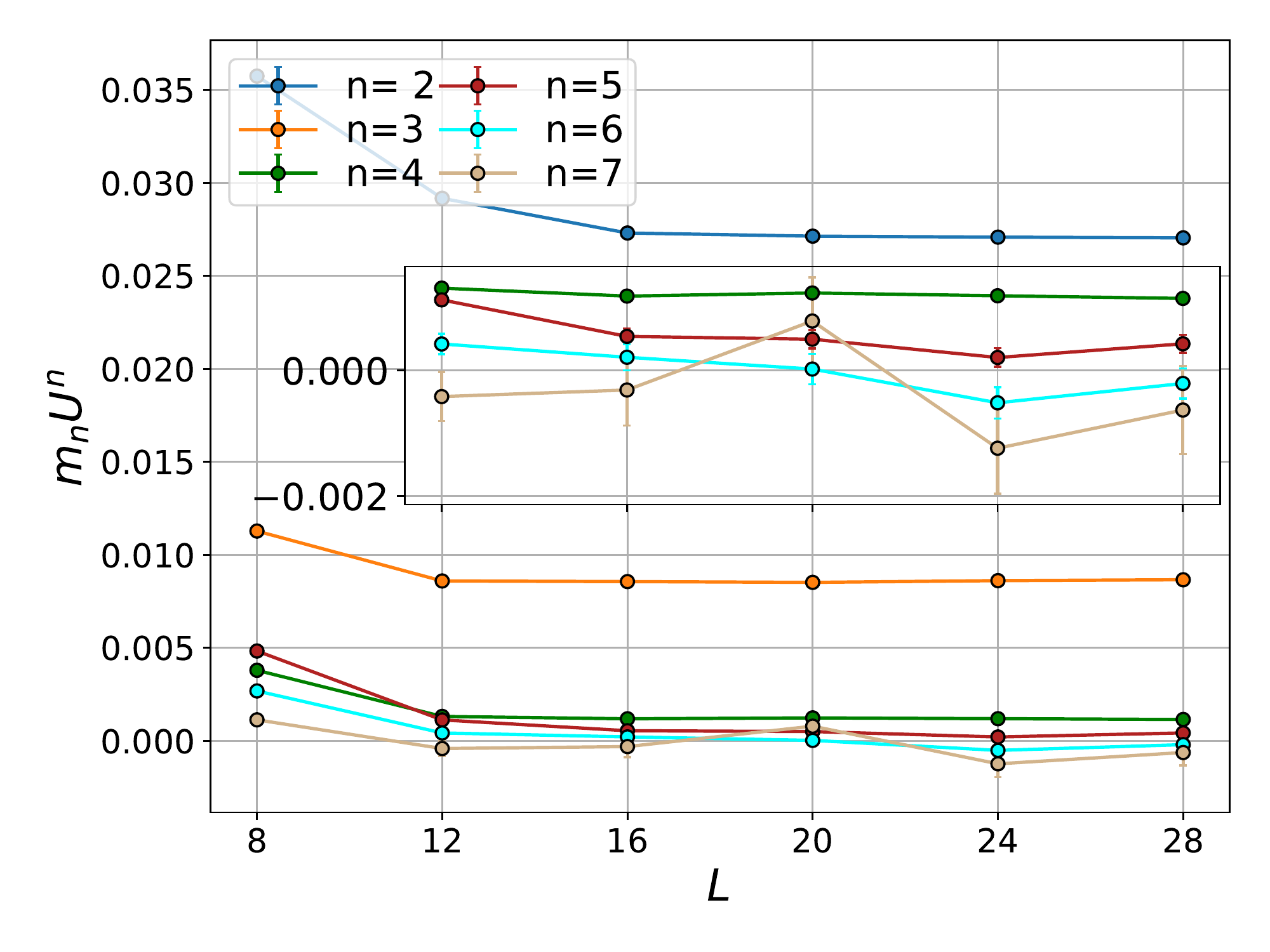} 
\caption{System-size study : Contribution to the magnetization of its expansion coefficients $m_{n}$ at different orders $n$, as a function of the linear system size $L$. The parameters of the perturbation expansion are $T=0.18$, $U=4$, $\alpha=0.1$~. The inset is the same plot for smaller contributions.}
\label{Sys_size_coeff}
\end{figure}

\begin{figure}[H]
\centering
\includegraphics[width=0.49\textwidth]{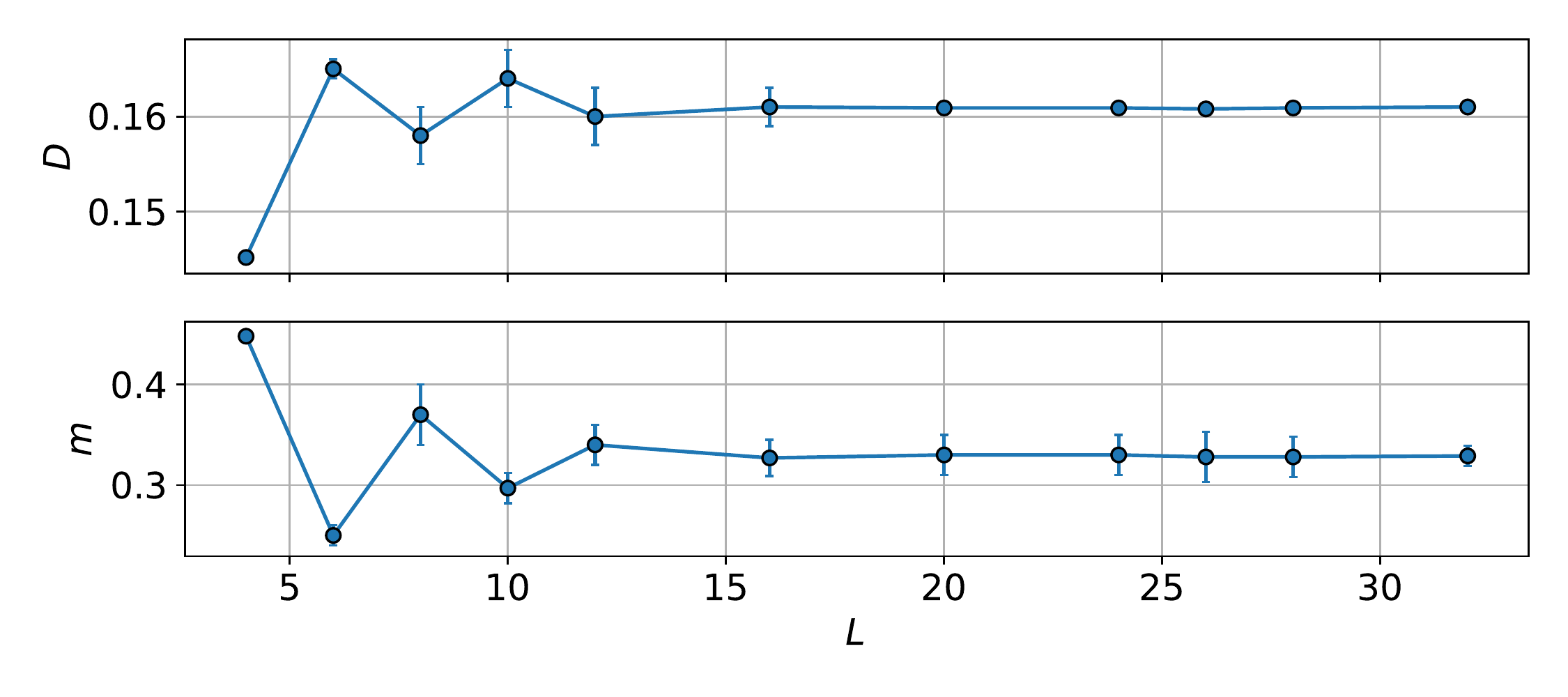} 
\caption{System-size study : Double-occupancy and magnetization at $U=4$, $T=0.15$ as a function of the linear system size $L$ for a cubic lattice of $L^{3}$ sites. }
\label{Sys_size}
\end{figure}

The diagrammatic expansion is formulated directly in the thermodynamic limit. In practice we solve the unperturbed Hamiltonian and tabulate for  finite $L$ the non-interacting Green's functions to build the diagrammatic expansion of the problem. Considering an expansion series of the magnetization at $U=4$, $T=0.18$ close to the phase transition ($T_{N}\simeq 0.19$ at an interaction $U=4$), we show in Fig.~\ref{Sys_size_coeff} that, up to relatively high order and negligible contribution to the resummed result, the contribution from each order is converged within statistical error for a linear system size $L\geq 20$. This shows that the final error on our results is dominated by statistical and resummation error. Finite size effects would only affect very high perturbation orders which can be neglected to determine the converging value of the series. This property remains valid when getting closer to the Néel temperature (see ~\cite{Spada}). In our study, we fix the system-size to $L^{3}=20^{3}$ sites on a cubic lattice which proves to be large enough to avoid finite-size effects. To illustrate this we show the evolution of the value of different observables with linear system size in Fig.~\ref{Sys_size}, with a convergence reached for $L\geq20$. 

\

\section{Grand-potential computations} 

\

\begin{figure}[H]
\centering
\includegraphics[width=0.5\textwidth]{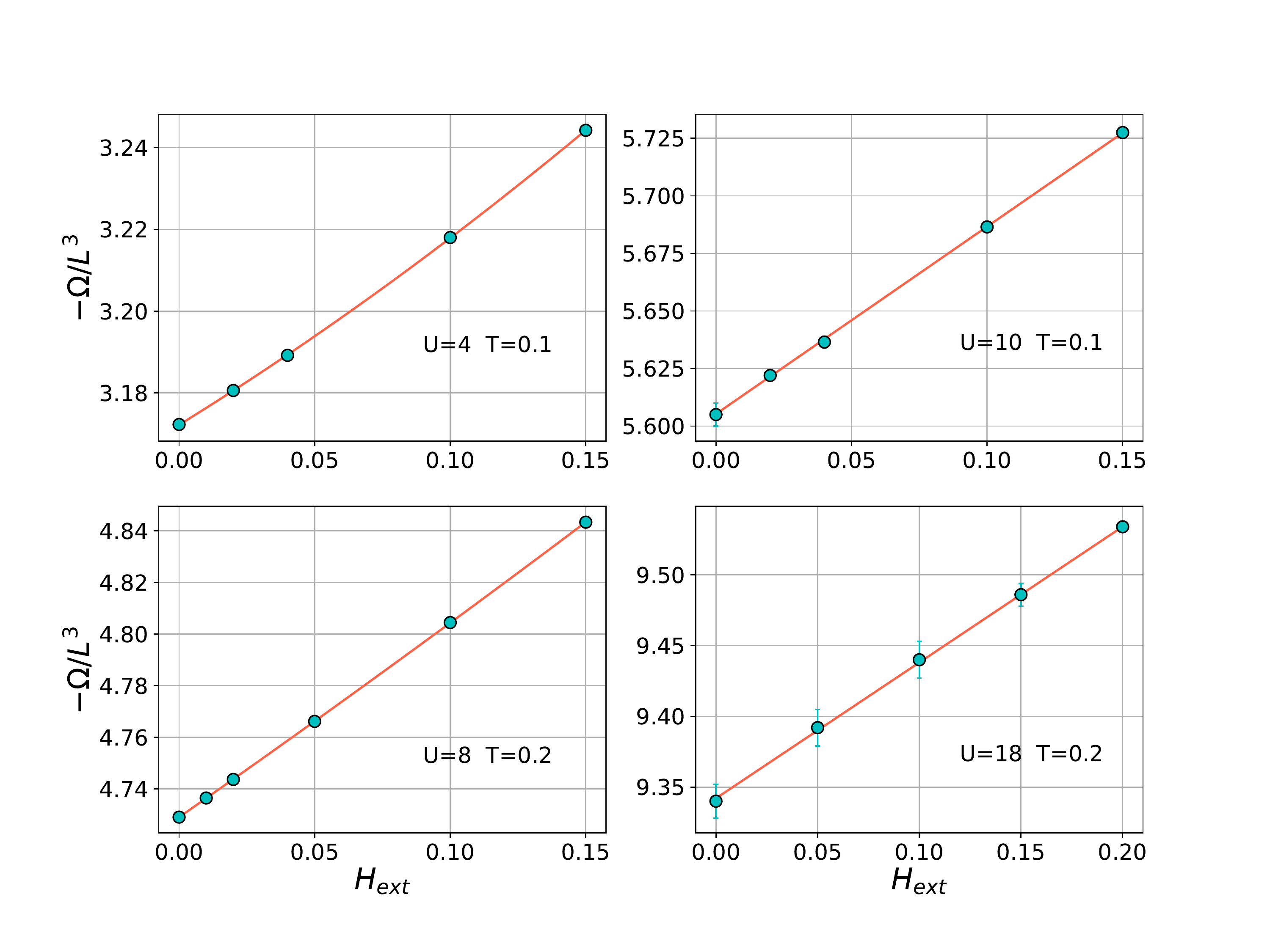}
\caption{Grand potential density $-\Omega/L^{3}$ as a function of the external staggered field $H_\mathrm{ext}$ for different values of the temperature and interaction parameters. The orange line corresponds to the best fit by a second (first) order polynomial on the top and bottom left (right) panels}
\label{P_h}
\end{figure}

Grand potential computations are shown in Fig. \ref{P_h}. The corresponding magnetization is obtained with a first or second order polynomial fit of the grand potential data as a function of the external field $H_{ext}$. This method for estimating indirectly the magnetization as a derivative of the grand potential proves to be efficient deep inside the antiferromagnetic dome, in the regime where magnetization is saturated to its zero temperature value. In the vicinity of the phase transition we expect the grand potential curve as a function of the external field $H_\mathrm{ext}$ to be flattening close to the origin $H_\mathrm{ext}=0$ as the temperature gets higher. These changes at very small fields are hard to resolve and we rely on direct computations of the magnetization in this regime.

\begin{table}[H]
    \centering
    \begin{tabular}{l|c|r}
    \hline
  T & U & $\alpha$ \\
  \hline
0.1 & 3 & (0.13,1) \\
0.2 & 3 & (0.01,1) \\
0.3 & 3 & (0.01,0.5) \\
0.05 & 4 & (0.3,1) \\
0.075 & 4 & (0.23) \\
0.1 & 4 & (0.25,0.4,0.75) \\
0.125 & 4 & (0.2,0.3,0.4,0.7) \\
0.15 & 4 & (0.25,0.55,0.7) \\
0.16 & 4 & (0.2,0.23,0.27) \\
0.17 & 4 & (0.12,0.13,0.19) \\
0.18 & 4 & (0.11,0.07) \\
0.19 & 4 & (0.02) \\
0.195 & 4 & (0,0.01) \\
0.2 & 4 & (0.01,0.05) \\
0.21 & 4 & (0.0001) \\
0.22 & 4 & (0.0001) \\
0.225 & 4 & (0.01,0.05) \\
0.23 & 4 & (0.0001) \\
0.24 & 4 & (0.0001) \\
0.25 & 4 & (0.0001,0.05) \\
0.275 & 4 & (0.01,0.05) \\
0.3 & 4 & (0.0001,0.01,1) \\
0.325 & 4 & (0.01) \\
0.35 & 4 & (0.01,1) \\
0.4 & 4 & (0.0001,1) \\
0.45 & 4 & (0.0001,1) \\
0.5 & 4 & (0.0001,1) \\
0.1 & 5 & (0.2,0.4,0.8) \\
0.2 & 5 & (0.2,0.4,0.5,0.8) \\
0.3 & 5 & (0.1,0.5,1) \\
0.1 & 6 & (0.4,0.8,1) \\
0.15 & 6 & (0.75,1) \\
0.2 & 6 & (0.45,0.83) \\
0.225 & 6 & (0.65,1) \\
0.25 & 6 & (0.6,1) \\
0.3 & 6 & (0.5,1) \\
0.325 & 6 & (0.05,0.5,0.85) \\
0.35 & 6 & (0.05,0.5,0.85) \\
0.375 & 6 & (0.05,0.5,0.85) \\
0.4 & 6 & (0.05,0.5,0.85) \\
  0.1 & 7 & (0.1,0.4,0.85) \\

    \end{tabular}
\begin{tabular}{l|c|r}
    \hline
  T & U & $\alpha$ \\
  \hline
0.2 & 7 & (0.25,0.45,0.85) \\
0.3 & 7 & (0.5,1) \\
  0.1 & 8 & (0.4,0.45,0.9) \\
0.15 & 8 & (0.2,0.3,0.5) \\
0.2 & 8 & (0.45,1) \\
0.25 & 8 & (0.1,0.45,1) \\
0.3 & 8 & (0.5,1) \\
0.325 & 8 & (0.5,0.8) \\
0.35 & 8 & (0.05,0.25,0.5) \\
0.375 & 8 & (0.05,0.25,0.5) \\
0.4 & 8 & (0.05,0.25,0.5) \\
0.1 & 9 & (0.5,0.8) \\
0.2 & 9 & (0.5,1) \\
0.3 & 9 & (0.5,1) \\
0.1 & 10 & (0.4,0.5,1) \\
0.2 & 10 & (0.5,1) \\
0.3 & 10 & (0.5,1) \\
0.1 & 11 & (0.5,1) \\
0.2 & 11 & (0.5,1) \\
0.3 & 11 & (0.5,1) \\
0.1 & 12 & (0.5,1) \\
0.2 & 12 & (0.5,1) \\
0.3 & 12 & (0.5,1) \\
0.1 & 13 & (0.5,1) \\
0.2 & 13 & (0.5,1) \\
0.3 & 13 & (0.5,1) \\
0.1 & 14 & (0.5,1) \\
0.2 & 14 & (0.5,1) \\
0.3 & 14 & (0.5,1) \\
0.1 & 15 & (0.5,1) \\
0.2 & 15 & (0.5,1) \\
0.3 & 15 & (0.5,1) \\
0.1 & 16 & (0.5,1) \\
0.2 & 16 & (0.5,1) \\
0.3 & 16 & (0.5,1) \\
0.1 & 17 & (0.5,1) \\
0.2 & 17 & (0.5,1) \\
0.3 & 17 & (0.5,1) \\
0.1 & 18 & (0.5,1) \\
0.2 & 18 & (0.5,1) \\  
0.3 & 18 & (0.5,1) \\
\end{tabular}
    \caption{List of the shift parameters $\alpha$ used to obtain the double-occupancy for different parameters $T$ and $U$. }
    \label{tab:D_alpha}

\end{table}

\begin{table}[H]
\centering
    \begin{tabular}{l|c|r}
    \hline
 T & U & $\alpha$ \\
  \hline
0.0175 & 2 & (0.25,0.3) \\
0.02 & 2 & (0.25,0.3) \\
0.0225 & 2 & (0.23,0.24,0.25) \\
0.025 & 2 & (0.2,0.23) \\
0.0275 & 2 & (0.23,0.24) \\
0.03 & 2 & (0.22,0.225) \\
0.0325 & 2 & (0.21,0.22,0.23) \\
0.035 & 2 & (0.19,0.2) \\
0.0375 & 2 & (0.15,0.16,0.17) \\
0.04 & 2 & (0.1,0.11,0.12) \\
0.0425 & 2 & (0.01,0.05) \\
0.1 & 3 & (0.12,0.125,0.13) \\
0.1 & 3.5 & (0.18,0.17) \\
0.075 & 4 & (0.18,0.19) \\
0.1 & 4 & (0.171,0.175,0.18) \\
0.125 & 4 & (0.167,0.17,0.2) \\
0.14 & 4 & (0.161) \\
0.15 & 4  & (0.157,0.162,0.17,0.25) \\
0.16 & 4 & (0.14,0.148,0.15,0.2) \\
0.165 & 4 & (0.138,0.139) \\
0.17 & 4 & (0.12,0.13,0.19) \\
0.175 & 4 & (0.1,0.11,0.12,0.13) \\
0.18 & 4 & (0.05,0.07,0.08,0.1) \\
0.183 & 4 & (0.078,0.082,0.084) \\
0.185 & 4 & (0.05,0.07) \\
0.187 & 4 & (0.053,0.057,0.061) \\
0.188 & 4 & (0.05,0.051,0.0513,0.052) \\
0.189 & 4 & (0.033,0.042,0.045) \\
0.19 & 4 & (0.02,0.025) \\
0.195 & 4 & (0.001,0.01) \\
0.2 & 4 & (0.01,0.05) \\
0.225 & 4 & (0.01,0.05) \\
0.25 & 4 & (0.0001,0.05) \\
0.1 & 5 & (0.14,0.17,0.22) \\
0.2 & 5 & (0.16,0.25) \\
0.1 & 6 & (0.7,0.8,1) \\
0.15 & 6 & (0.7,1) \\
0.2 & 6 & (0.7,1) \\
0.22 & 6 & (0.7,1) \\
0.24 & 6 & (0.18,0.22) \\
0.25 & 6 & (0.2,0.25) \\
0.26 & 6 & (0.1,0.12) \\
0.27 & 6 & (0.19,0.18) \\
0.28 & 6 & (0.17,0.173,0.177,0.18) \\
0.29 & 6 & (0.17,0.175) \\
0.295 & 6 & (0.162,0.17) \\
0.3 & 6 & (0.14,0.15) \\
0.305 & 6 & (0.13,0.14) \\
0.31 & 6 & (0.13,0.145) \\
0.315 & 6 & (0.01,0.02) \\
0.32 & 6 & (0.01) \\
0.1 & 7 & (0.4,0.65,0.8) \\
\end{tabular}
    \caption{List of the shift parameters $\alpha$ used to obtain the magnetization for different parameters $T$ and $U$. }
    \label{tab:m_alpha}

\end{table}

\begin{table}[H]
\scriptsize
\centering
    \begin{tabular}{l|c|c|r}
    \hline
  T & U & $H_\mathrm{ext}$ & $\alpha$ \\
  \hline
0.1 & 4 & 0. & (0.7,1) \\
0.1 & 4 & 0.02 & (0.7,1) \\
0.1 & 4 & 0.04 & (0.7,1) \\
0.1 & 4 & 0.1 & (0.7,1) \\
0.1 & 4 & 0.15 & (0.7,1) \\
0.11 & 4 & 0. & (0.6,0.7,1) \\
0.12 & 4 & 0. & (0.6,0.7,1) \\
0.13 & 4 & 0. & (0.6,0.7,1) \\
0.14 & 4  & 0. & (0.6,0.7,1) \\
0.15 & 4 & 0. & (0.05,0.5,0.7,1) \\
0.15 & 4 & 0.01 & (0.05,0.5,0.7,1) \\
0.15 & 4 & 0.02 & (0.05,0.5,0.7,1) \\
0.16 & 4 & 0. & (0.05,0.5) \\
0.25 & 4 & 0. & (0.05,0.5,0.7) \\
0.3 & 4 & 0. & (0.05,0.5,0.7) \\
0.35 & 4 & 0. & (0.05,0.5,0.7) \\
0.4 & 4 & 0. & (0.05,0.5,1) \\
0.45 & 4 & 0. & (0.05,0.5,1) \\
0.5 & 4 & 0 & (0.05,0.5,1) \\
0.75 & 4 & 0 & (0.05,0.5,1) \\
0.1 & 5 & 0. & (0.7,1) \\
0.1 & 5 & 0.02 & (0.7,1) \\
0.1 & 5 & 0.04 & (0.7,1) \\
0.1 & 5 & 0.1 & (0.7,1) \\
0.1 & 5 & 0.15 & (0.7,1) \\
0.2 & 5 & 0 & (0.7,1) \\
0.2 & 5 & 0.02 & (0.7,1) \\
0.2 & 5 & 0.04 & (0.7,1) \\
0.2 & 5 & 0.1 & (0.7,1) \\
0.2 & 5 & 0.15 & (0.7,1) \\
0.1 & 6 & 0. & (0.7,1) \\
0.1 & 6 & 0.02 & (0.7,1) \\
0.1 & 6 & 0.05 & (0.7,1) \\
0.1 & 6 & 0.1 & (0.7,1) \\
0.1 & 6 & 0.15 & (0.7,1) \\
0.2 & 6 & 0. & (0.7,1) \\
0.2 & 6 & 0.005 & (0.7,1) \\
0.2 & 6 & 0.01 & (0.7,1) \\
0.2 & 6 & 0.015 & (0.7,1) \\
0.2 & 6 & 0.02 & (0.7,1) \\
0.2 & 6 & 0.025 & (0.7,1) \\
0.1 & 7 & 0 & (0.7,1) \\
0.1 & 7 & 0.02 & (0.7,1) \\
0.1 & 7 & 0.04 & (0.7,1) \\
0.1 & 7 & 0.1 & (0.7,1) \\
0.1 & 7 & 0.15 & (0.7,1) \\
0.2 & 7 & 0 & (0.7,1) \\
0.2 & 7 & 0.02 & (0.7,1) \\
0.2 & 7 & 0.04 & (0.7,1) \\
0.2 & 7 & 0.1 & (0.7,1) \\
0.2 & 7 & 0.15 & (0.7,1) \\
0.1 & 8 & 0 & (0.7,1) \\
0.1 & 8 & 0.02 & (0.7,1) \\
0.1 & 8 & 0.04 & (0.7,1) \\
0.1 & 8 & 0.1 & (0.7,1) \\
0.1 & 8 & 0.15 & (0.7,1) \\
0.2 & 8 & 0. & (0.7,1) \\
0.2 & 8 & 0.01 & (0.7,1) \\
0.2 & 8 & 0.02 & (0.7,1) \\
0.2 & 8 & 0.05 & (0.7,1) \\
0.2 & 8 & 0.1 & (0.7,1) \\
0.2 & 8 & 0.15 & (0.7,1) \\
0.1 & 9 & 0 & (0.7,1) \\
0.1 & 9 & 0.02 & (0.7,1) \\
0.1 & 9 & 0.04 & (0.7,1) \\
0.1 & 9 & 0.1 & (0.7,1) \\
0.1 & 9 & 0.15 & (0.7,1) \\
0.2 & 9 & 0 & (0.7,1) \\
0.2 & 9 & 0.02 & (0.7,1) \\
0.2 & 9 & 0.04 & (0.7,1) \\
0.2 & 9 & 0.1 & (0.7,1) \\
0.2 & 9 & 0.15 & (0.7,1) \\
0.1 & 10 & 0 & (0.7,1) \\
0.1 & 10 & 0.02 & (0.7,1) \\
0.1 & 10 & 0.04 & (0.7,1) \\
0.1 & 10 & 0.1 & (0.7,1) \\
0.1 & 10 & 0.15 & (0.7,1) \\
0.2 & 10 & 0 & (0.7,1) \\
0.2 & 10 & 0.02 & (0.7,1) \\
0.2 & 10 & 0.04 & (0.7,1) \\
0.2 & 10 & 0.1 & (0.7,1) \\
    \end{tabular}
\begin{tabular}{l|c|c|r}
    \hline
  T & U & $H_\mathrm{ext}$ & $\alpha$ \\
  \hline

0.2 & 10 & 0.15 & (0.7,1) \\
0.1 & 11 & 0 & (0.7,1) \\
0.1 & 11 & 0.05 & (0.7,1) \\
0.1 & 11 & 0.1 & (0.7,1) \\
0.1 & 11 & 0.15 & (0.7,1) \\
0.1 & 11 & 0.2 & (0.7,1) \\
0.2 & 11 & 0 & (0.7,1) \\
0.2 & 11 & 0.05 & (0.7,1) \\
0.2 & 11 & 0.1 & (0.7,1) \\
0.2 & 11 & 0.15 & (0.7,1) \\
0.2 & 11 & 0.2 & (0.7,1) \\
0.1 & 12 & 0 & (0.7,1) \\
0.1 & 12 & 0.05 & (0.7,1) \\
0.1 & 12 & 0.1 & (0.7,1) \\
0.1 & 12 & 0.15 & (0.7,1) \\
0.1 & 12 & 0.2 & (0.7,1) \\
0.2 & 12 & 0 & (0.7,1) \\
0.2 & 12 & 0.05 & (0.7,1) \\
0.2 & 12 & 0.1 & (0.7,1) \\
0.2 & 12 & 0.15 & (0.7,1) \\
0.2 & 12 & 0.2 & (0.7,1) \\
0.1 & 13 & 0 & (0.7,1) \\
0.1 & 13 & 0.05 & (0.7,1) \\
0.1 & 13 & 0.1 & (0.7,1) \\
0.1 & 13 & 0.15 & (0.7,1) \\
0.1 & 13 & 0.2 & (0.7,1) \\
0.2 & 13 & 0 & (0.7,1) \\
0.2 & 13 & 0.05 & (0.7,1) \\
0.2 & 13 & 0.1 & (0.7,1) \\
0.2 & 13 & 0.15 & (0.7,1) \\
0.2 & 13 &  0.2 & (0.7,1) \\
0.1 & 14 & 0 & (0.7,1) \\
0.1 & 14 & 0.05 & (0.7,1) \\
0.1 & 14 & 0.1 & (0.7,1) \\
0.1 & 14 & 0.15 & (0.7,1) \\
0.1 & 14 & 0.2 & (0.7,1) \\
0.2 & 14 & 0 & (0.7,1) \\
0.2 & 14 & 0.05 & (0.7,1) \\
0.2 & 14 & 0.1 & (0.7,1) \\
0.2 & 14 & 0.15 & (0.7,1) \\
0.2 & 14 & 0.2 & (0.7,1) \\
0.1 & 15 & 0 & (0.7,1) \\
0.1 & 15 & 0.05 & (0.7,1) \\
0.1 & 15 & 0.1 & (0.7,1) \\
0.1 & 15 & 0.15 & (0.7,1) \\
0.1 & 15 & 0.2 & (0.7,1) \\
0.2 & 15 & 0 & (0.7,1) \\
0.2 & 15 & 0.05 & (0.7,1) \\
0.2 & 15 & 0.1 & (0.7,1) \\
0.2 & 15 & 0.15 & (0.7,1) \\
0.2 & 15 & 0.2 & (0.7,1) \\
0.1 & 16 & 0 & (0.7,1) \\
0.1 & 16 & 0.05 & (0.7,1) \\
0.1 & 16 & 0.1 & (0.7,1) \\
0.1 & 16 & 0.15 & (0.7,1) \\
0.1 & 16 & 0.2 & (0.7,1) \\
0.2 & 16 & 0 & (0.7,1) \\
0.2 & 16 & 0.05 & (0.7,1) \\
0.2 & 16 & 0.1 & (0.7,1) \\
0.2 & 16 & 0.15 & (0.7,1) \\
0.2 & 16 & 0.2 & (0.7,1) \\
0.1 & 17 & 0 & (0.7,1) \\
0.1 & 17 & 0.05 & (0.7,1) \\
0.1 & 17 & 0.1 & (0.7,1) \\
0.1 & 17 & 0.15 & (0.7,1) \\
0.1 & 17 & 0.2 & (0.7,1) \\
0.2 & 17 & 0 & (0.7,1) \\
0.2 & 17 & 0.05 & (0.7,1) \\
0.2 & 17 & 0.1 & (0.7,1) \\
0.2 & 17 & 0.15 & (0.7,1) \\
0.2 & 17 & 0.2 & (0.7,1) \\
0.1 & 18 & 0 & (0.7,1) \\
0.1 & 18 & 0.05 & (0.7,1) \\
0.1 & 18 & 0.1 & (0.7,1) \\
0.1 & 18 & 0.15 & (0.7,1) \\
0.1 & 18 & 0.2 & (0.7,1) \\
0.2 & 18 & 0 & (0.7,1) \\
0.2 & 18 & 0.05 & (0.7,1) \\
0.2 & 18 & 0.1 & (0.7,1) \\
0.2 & 18 & 0.15 & (0.7,1) \\
0.2 & 18 & 0.2 & (0.7,1) \\
\end{tabular}
    \caption{List of the shift parameters $\alpha$ used to obtain the grand potential density for different parameters $T$, $U$ and $H_\mathrm{ext}$. }
    \label{tab:P_alpha}

\end{table}

\bibliographystyle{ieeetr}
\bibliography{references.bib}